\documentclass[a4paper]{jpconf}
\usepackage{graphicx}
\begin{document}
\title{Pulsed Beam Tests at the SANAEM RFQ Beamline}

\author{G Turemen$^{1,2}$, Y Akgun$^{1}$, A Alacakir$^{1}$, I Kilic$^{1}$, B Yasatekin$^{1,2}$, E Ergenlik$^{3}$, S Ogur$^{3}$, E Sunar$^{3}$, V Yildiz$^{3}$, F Ahiska$^{4}$, E Cicek$^{5}$ and G Unel$^{6}$}

\address{$^1$Saraykoy Nuclear Research and Training Center, Turkish Atomic Energy Authority, Ankara, Turkey}
\address{$^2$Department of Physics, Ankara University, Ankara, Turkey}
\address{$^3$Department of Physics, Bogazici University, Istanbul, Turkey}
\address{$^4$EPROM Electronic Project \& Microwave Ind. and Trade Ltd. Co., Ankara, Turkey}
\address{$^5$Department of Physics, Gazi University, Ankara, Turkey}
\address{$^6$Department of Physics \& Astronomy, University of California at Irvine, Irvine, California, USA}

\ead{gorkem.turemen@ankara.edu.tr}

\begin{abstract}
A proton beamline consisting of an inductively coupled plasma (ICP) source, two solenoid magnets, two steerer magnets and a radio frequency quadrupole (RFQ) is developed at the Turkish Atomic Energy Authority's (TAEA) Saraykoy Nuclear Research and Training Center (SNRTC-SANAEM) in Ankara. In Q4 of 2016, the RFQ was installed in the beamline. The high power tests of the RF power supply and the RF transmission line were done successfully. The high power RF conditioning of the RFQ was performed recently. The 13.56 MHz ICP source was tested in two different conditions, CW and pulsed. The characterization of the proton beam was done with ACCTs, Faraday cups and a pepper-pot emittance meter. Beam transverse emittance was measured in between the two solenoids of the LEBT. The measured beam is then reconstructed at the entrance of the RFQ by using computer simulations to determine the optimum solenoid currents for acceptance matching of the beam. This paper will introduce the pulsed beam test results at the SANAEM RFQ beamline. In addition, the high power RF conditioning of the RFQ will be discussed.
\end{abstract}

\section{Introduction}
SANAEM is finalizing the construction of a proton beamline to be used for training and educational purposes. All of the components were designed and produced locally to train accelerator physicists and engineers on the job. The full proton beamline, which is designed for pulsed beam operation, composed of an ion source, a low energy beam transport and an RFQ operating at 352.21 MHz, together with the RF components was designed and installed. This note summarizes the ongoing commissioning work of both the proton beamline and the RF power supply unit (PSU).

\section{SANAEM RFQ Beamline}
\begin{figure}[h]
   \centering
   \includegraphics[width=35pc]{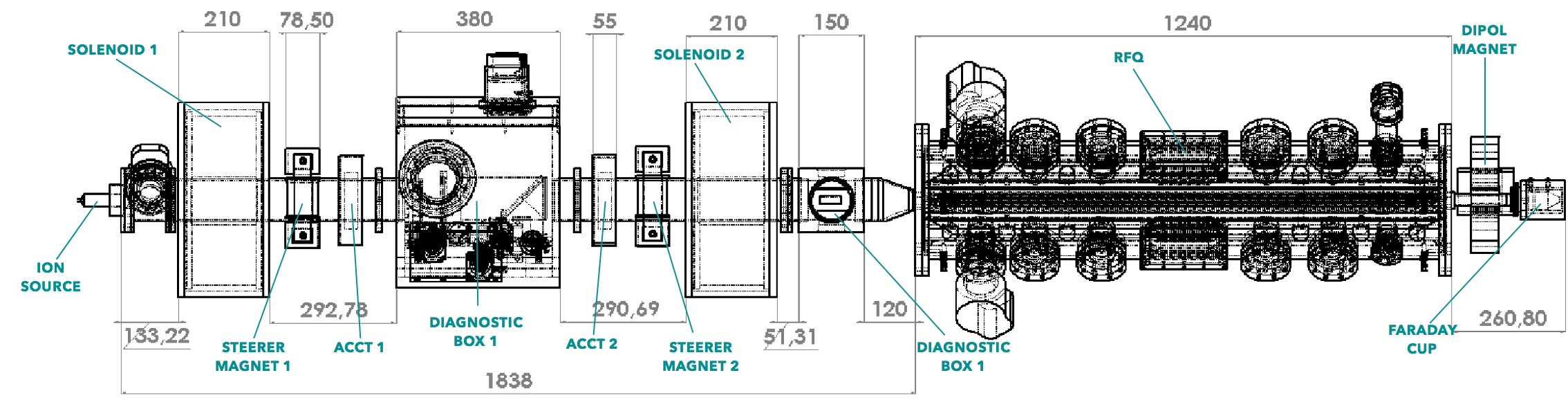}
   \includegraphics[width=35pc]{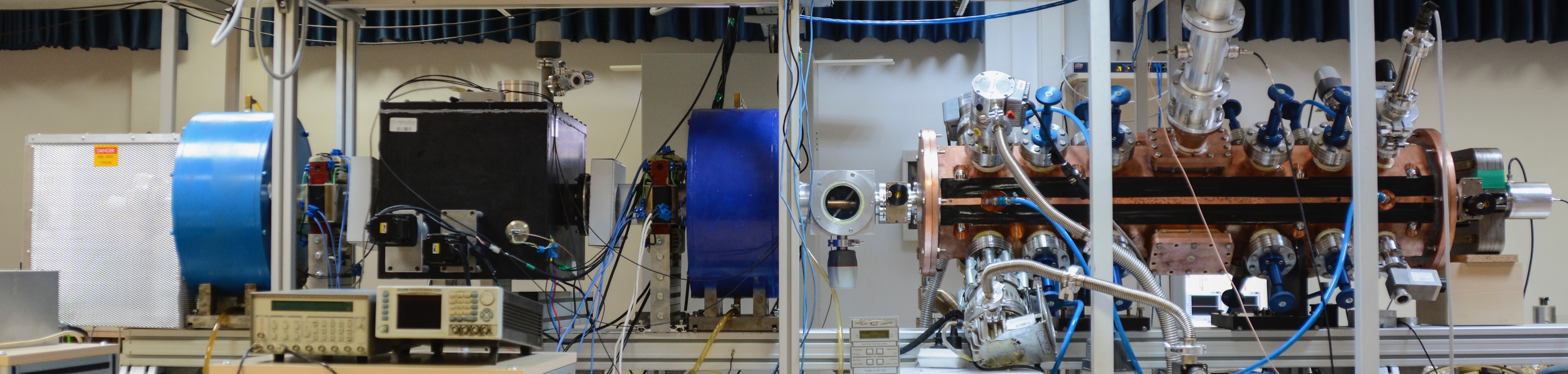}
   \caption{\label{line}SANAEM RFQ beamline.}
\end{figure}

The beamline of the SANAEM RFQ consists of an ion source, a LEBT (Low Energy Beam Transport) line, a 1.3 MeV RFQ and an energy spectrometer. The drawing and a view of the LEBT line and the RFQ cavity is shown in Fig.\ref{line}. The ion source and the LEBT line components were designed to match the input beam requirements of the RFQ. Since a pulsed RF operation is envisaged for the cavity, a pulsed proton beam is to be delivered to the RFQ.

\subsection{Pulsed Beam Extraction}
\begin{figure}[h]
\centering
\begin{minipage}{11pc}
\includegraphics[width=11pc]{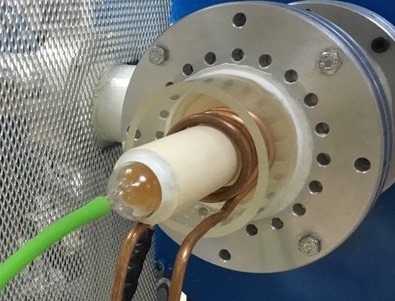}
\end{minipage}\hspace{1pc}
\begin{minipage}{11pc}
\includegraphics[width=11pc]{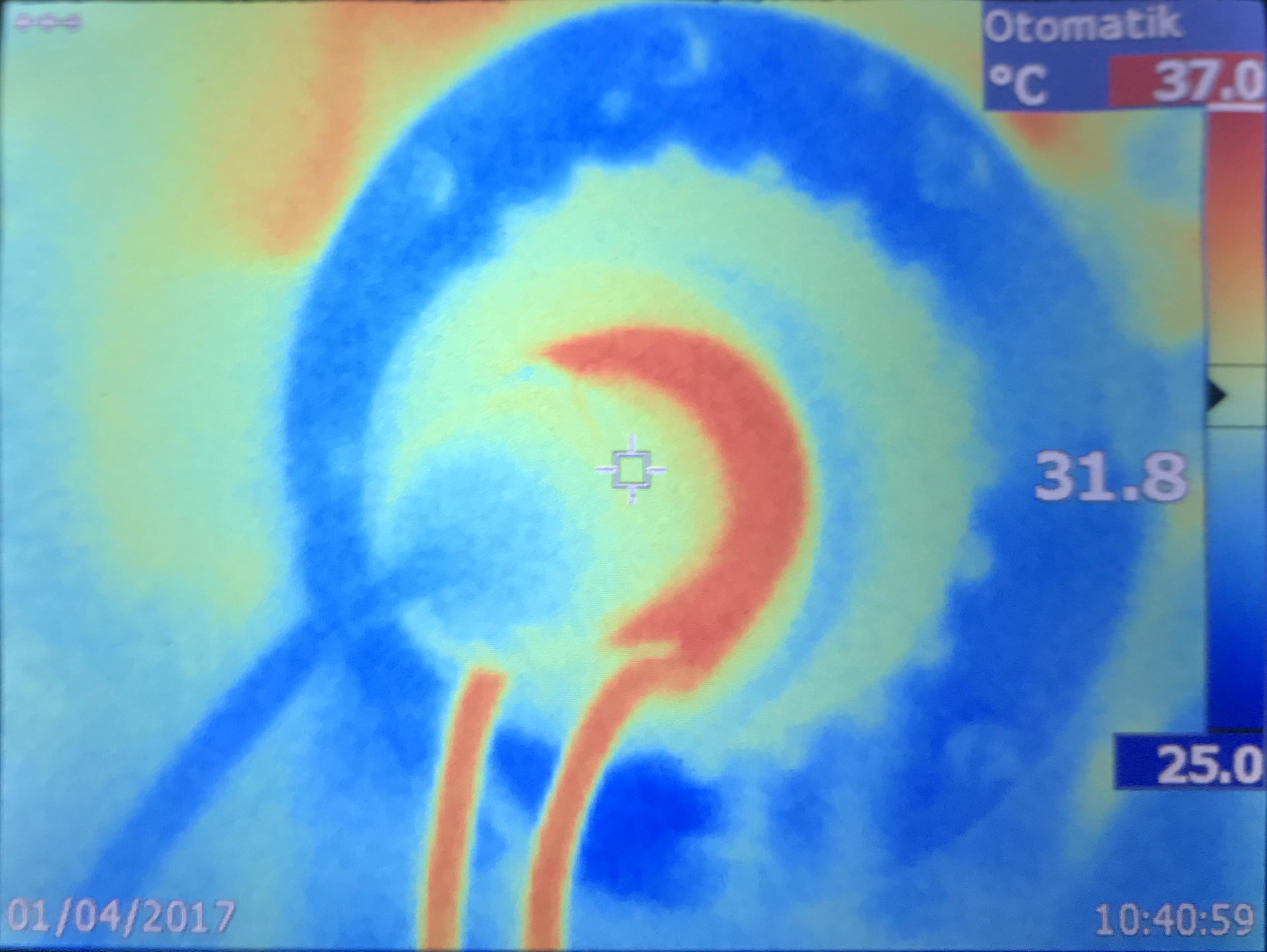}
\end{minipage}\hspace{1pc}
\begin{minipage}{11pc}
\includegraphics[width=11pc]{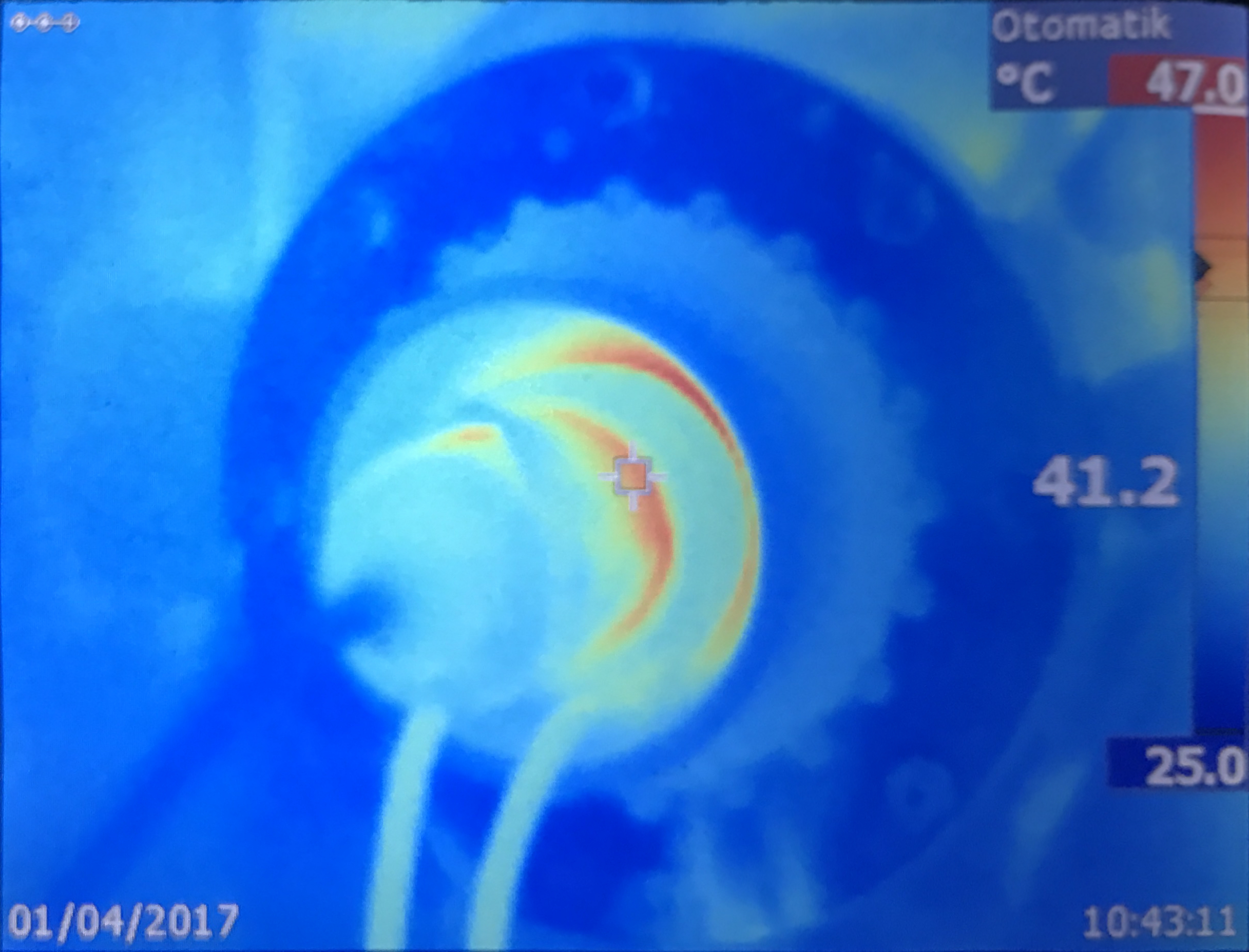}
\end{minipage}
\caption{\label{ikfoto2}The measured temperatures of the plasma chamber (left) with pulsed (middle) and continuous (right) RF power.}
\end{figure}

A 1 kW, 13.56 MHz ICP ion source was developed to generate a 20 keV proton beam with a current of several hundreds of micro-amperes. The ICP source consists of a plasma chamber of 28 mm diameter, 2.5 turns RF antenna, and two electrode extraction system (Fig.\ref{ikfoto2}). 

With the current configuration, the pulsed proton beam can be produced with two different pulsing approaches. One of these, as in a previously introduced study \cite{ipac15s}, is to use a pulsed high voltage to extract the protons from the plasma. In this approach the plasma is continuous and a PFN (pulse forming network) is used to pulse the high voltage electrode of the extraction region. The other approach is pulsing the RF power, consequently the plasma, and supply a DC high voltage to the extraction electrode. In this configuration, the pulse length of the extracted beam is specified by the duty factor of the RF source. Since a low duty factor (1\%) is used to excite the plasma, the temperature of the plasma chamber does not increase significantly and hence no water cooling is needed  for the chamber. 
The temperature of the alumina plasma chamber and the copper RF antenna is measured with an infrared temperature sensor for these two different approaches (Fig.\ref{ikfoto2}) for 60 seconds operation time. Consequently, the second approach was found to be more convenient. On the extraction side, a plasma electrode with a 3 mm hole diameter was used to keep  the transverse emittance low. The developed ICP ion source is suitable for long operation times with 1 Hz, 10 ms RF pulses.

\section{Proton Beam Characterization}
SANAEM RFQ's LEBT line has two diagnostic boxes and two ACCTs for beam characterization. One of the diagnostic boxes (Box-1) is located in between the LEBT solenoids (SOL-1 and SOL-2) and the other (Box-2) is located just before the RFQ cavity. The Box-2 has a Faraday cup and a phosphor screen and the Box-1 has an additional tool, a pepper-pot plate, for measuring the emittance of the beam. A CMOS camera with 85 mm-f/2.8 lens was used for the profile and emittance measurements. All the diagnostic components were developed locally, except the phosphor screen (P43) in the Box-1. Box-2 will also serve to house a beam chopper which is under production phase.

\subsection{Beam Size Measurements}
\begin{figure}[h]
\centering
\begin{minipage}{11pc}
\includegraphics[width=11pc]{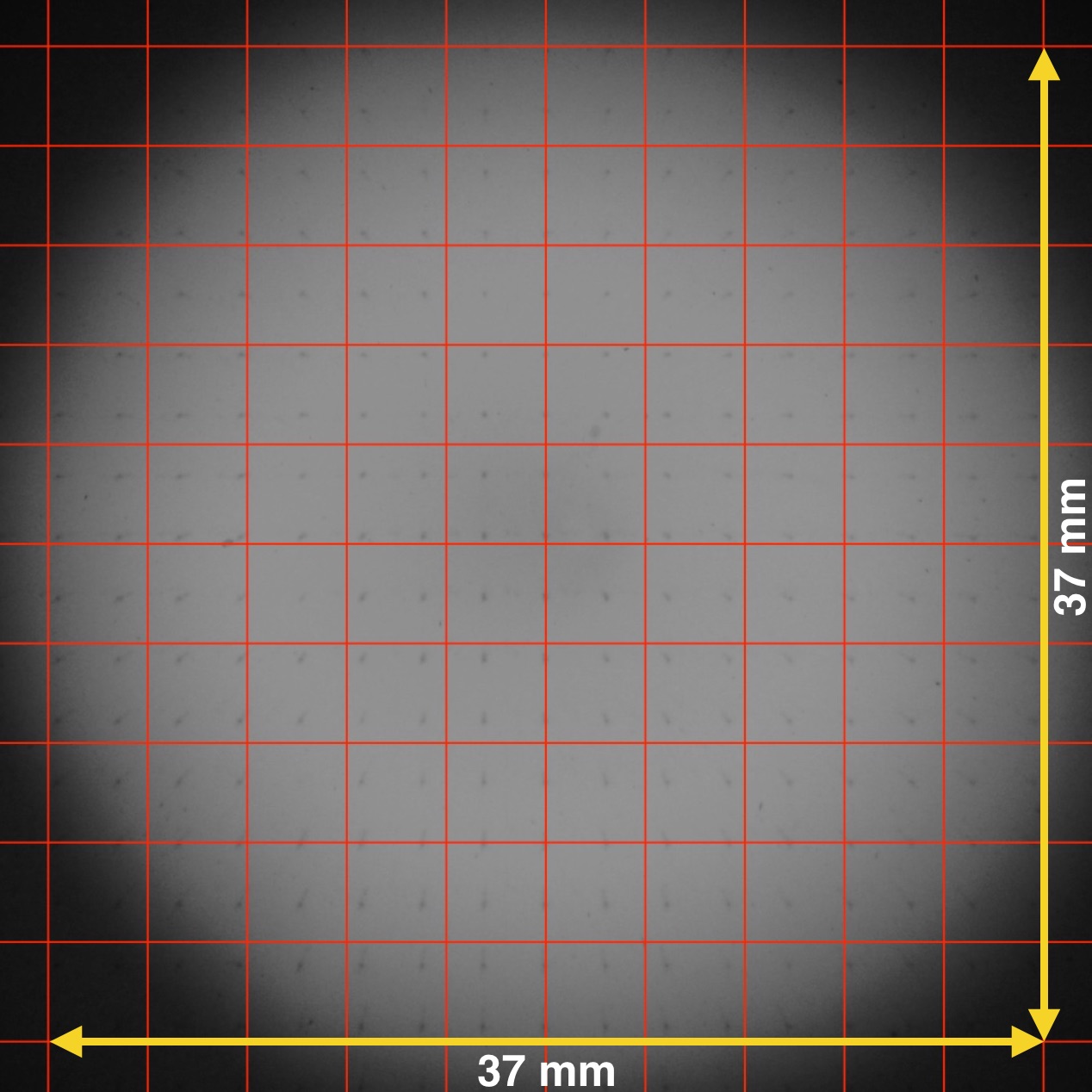}
\end{minipage}\hspace{1pc}
\begin{minipage}{16pc}
\includegraphics[width=16pc]{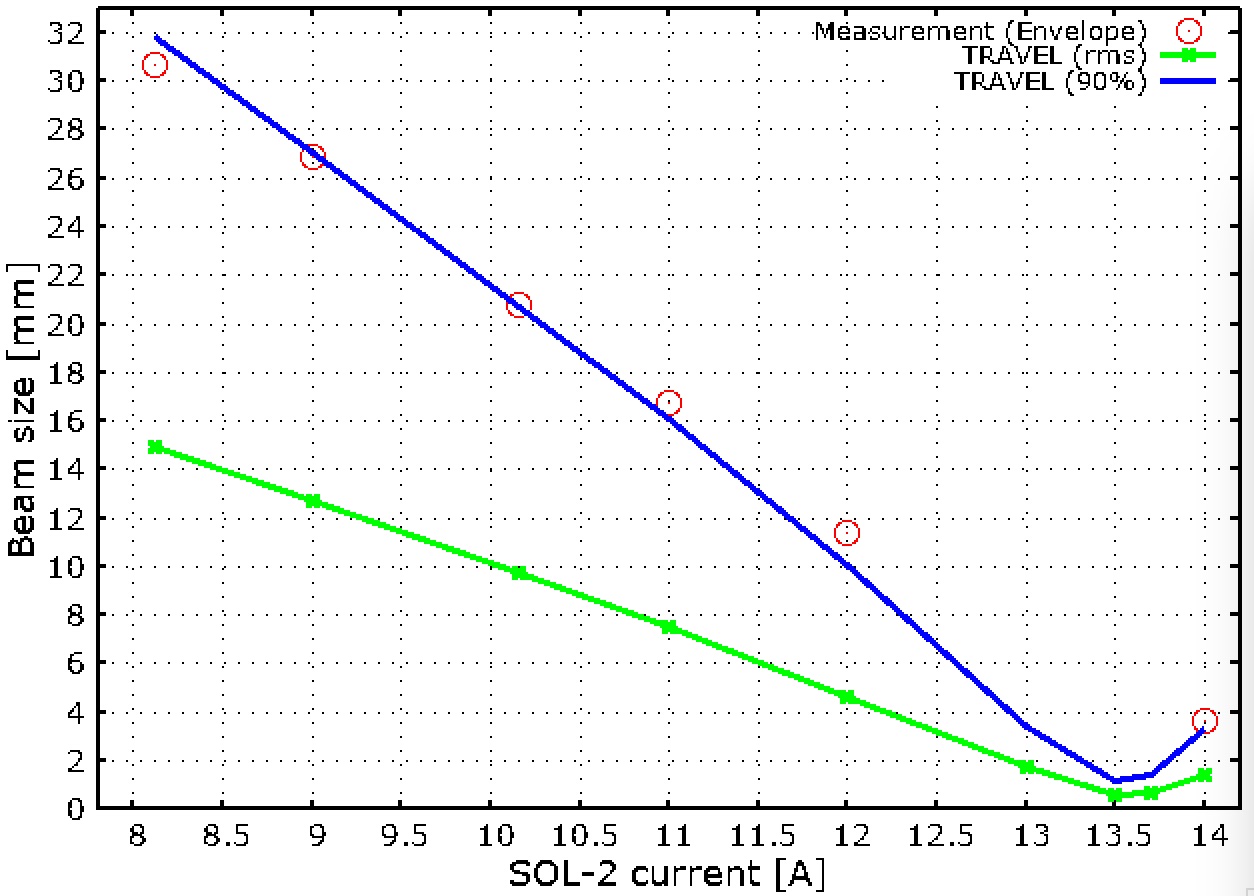}
\end{minipage}
\caption{\label{box12s}Measured beam spot in Box-1 (left) and simulated and measured beam spot sizes in the Box-2 (right).}
\end{figure}

The beam size measurements are performed to ensure full transmission of the beam throughout the LEBT line (with a beam pipe diameter of 90 mm). The spot size of the extracted beam was measured in Box-1 and Box-2. A number of measurements were performed to characterize the beam behavior in different operating conditions, RF power and solenoid currents. The measured beam spot at Box-1 is shown in Fig.\ref{box12s} for 200 W RF power and 13 A SOL-1 current. Similarly, the beam spot size was measured just before the RFQ cavity (in Box-2) for 200 W RF power with a SOL-1 current of 13 A and various SOL-2 currents. As shown in Fig.\ref{box12s}, the SOL-2 is able to over-focus the beam on the phosphor screen in Box-2 with a current of 13.7 A.

However, there is a space of 20 cm between the screen in the Box-2 and the RFQ plate, the required beam size (2.8 mm (100\%) at the RFQ entrance) is provided by using the solenoids. In this case, the beam spot size always remains smaller than the beam pipe diameter thus full transmission of the low current beam is achieved.

\subsection{Emittance Measurements}
\begin{figure}[h]
\centering
\begin{minipage}{10pc}
\includegraphics[width=10pc]{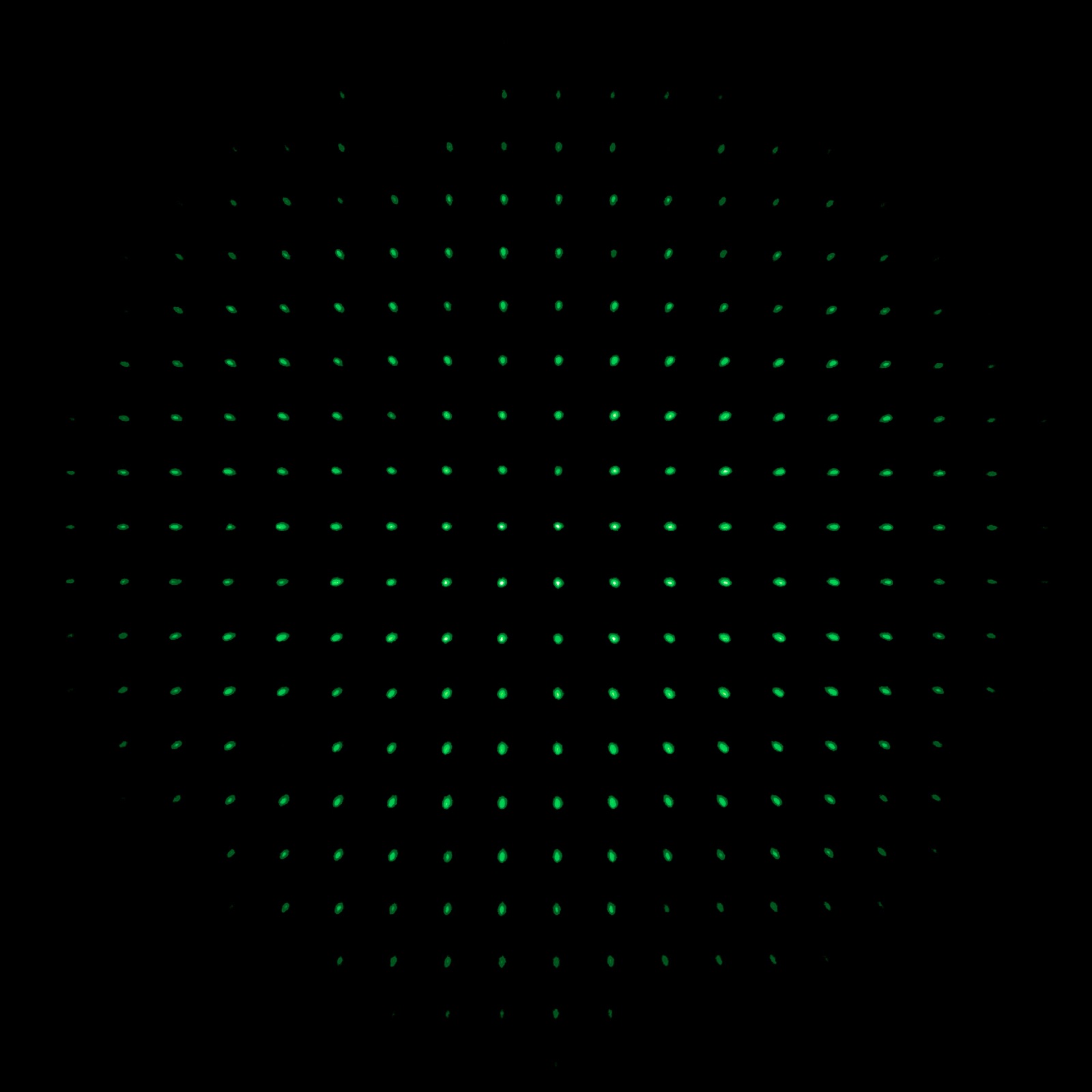}
\end{minipage}\hspace{1pc}
\begin{minipage}{12.5pc}
\includegraphics[width=12.5pc]{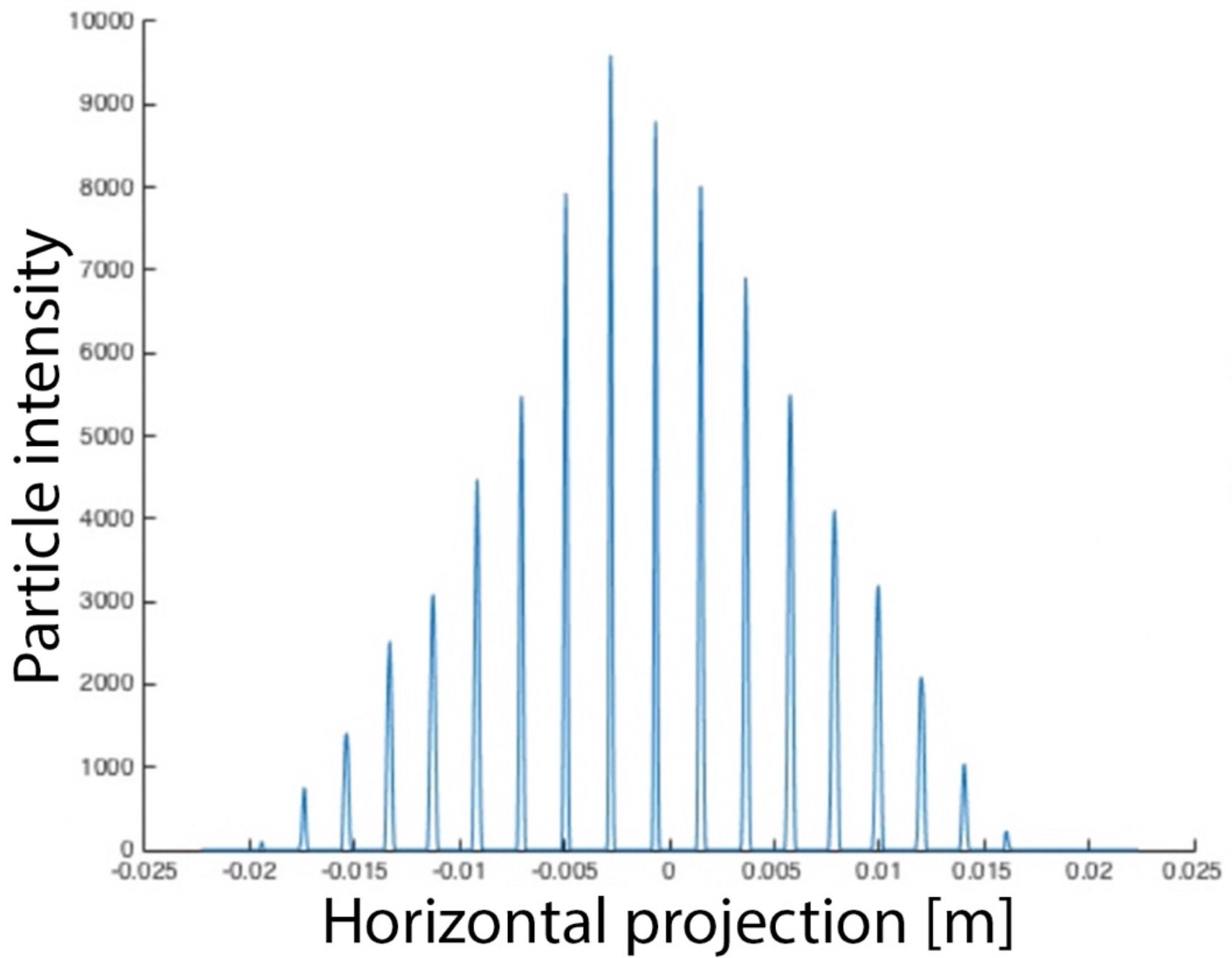}
\end{minipage}\hspace{1pc}
\begin{minipage}{12.5pc}
\includegraphics[width=12.5pc]{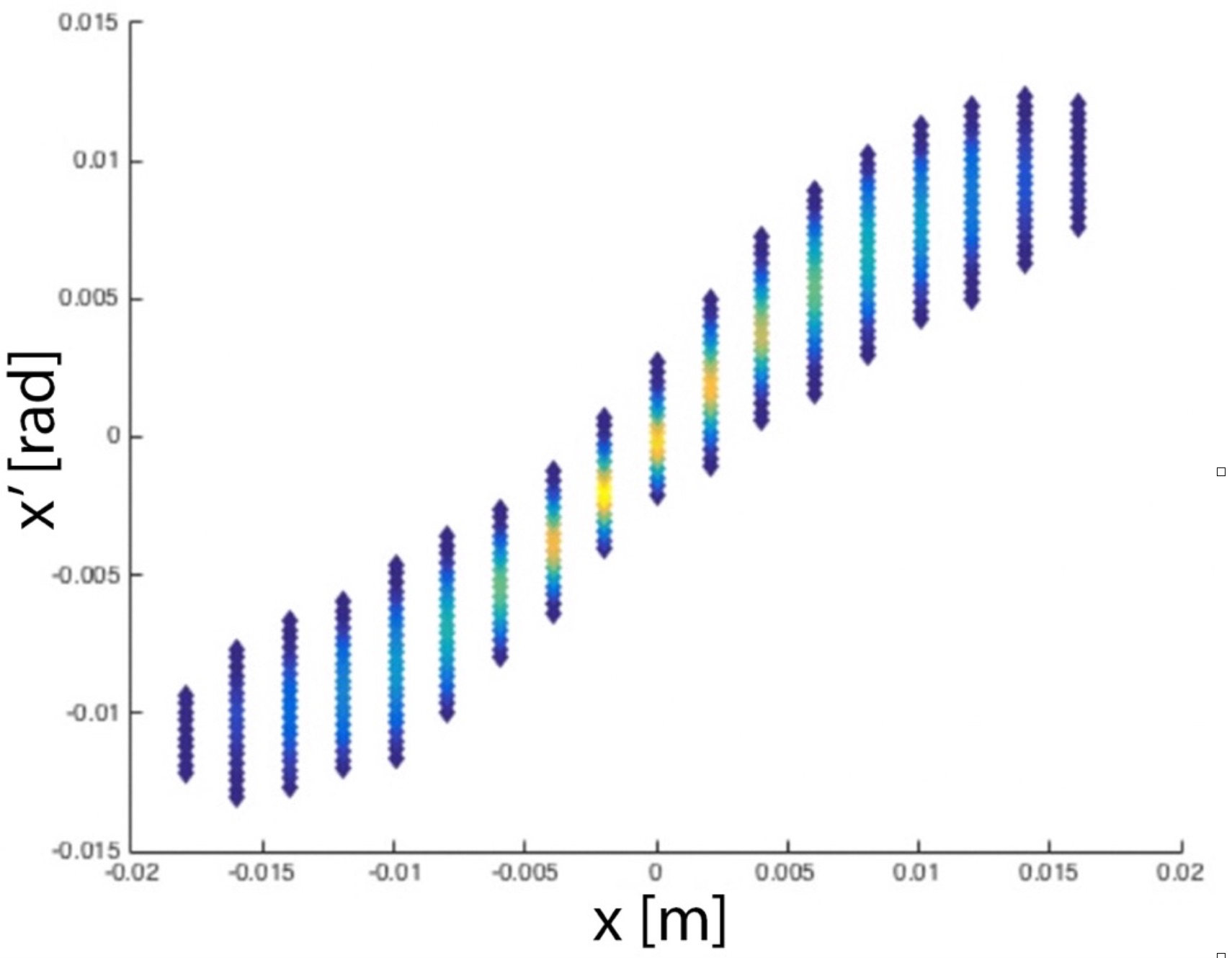}
\end{minipage}
\caption{\label{ppfinal}Captured image of the beam on the P43 screen (left), pepper-pot profile (middle) and transverse phase space (rigth) of the measured beam.}
\end{figure}

The characteristics of the transverse emittance measurement system with a pepper-pot method at Box-1 was presented in a previous study \cite{karakutu}. A 20 keV 10 ms proton pulse with a repetition frequency of 1 Hz was passed over the pepper-pot plate and its profile was captured with the camera on P43 screen (76 mm downstream). The captured pepper-pot image (Fig.\ref{ppfinal} on the left) was examined with a locally developed software and the transverse emittance (normalized rms) of the beam was calculated as 0.058 $\mp$0.001 and 0.050 $\mp$0.001 ${\mu}$m for vertical and horizontal axes, respectively. Although the whole system is cylindrical symmetric, the possible reasons for the difference between the measured emittance values are aberrations in the solenoid fields, measurement and calculation errors. As a future study, solenoid scan method will be considered to compare with the emittance values obtained with the pepper-pot.

\subsection{Current Measurements}
\begin{figure}[h]
   \centering
   \includegraphics[width=26.5pc]{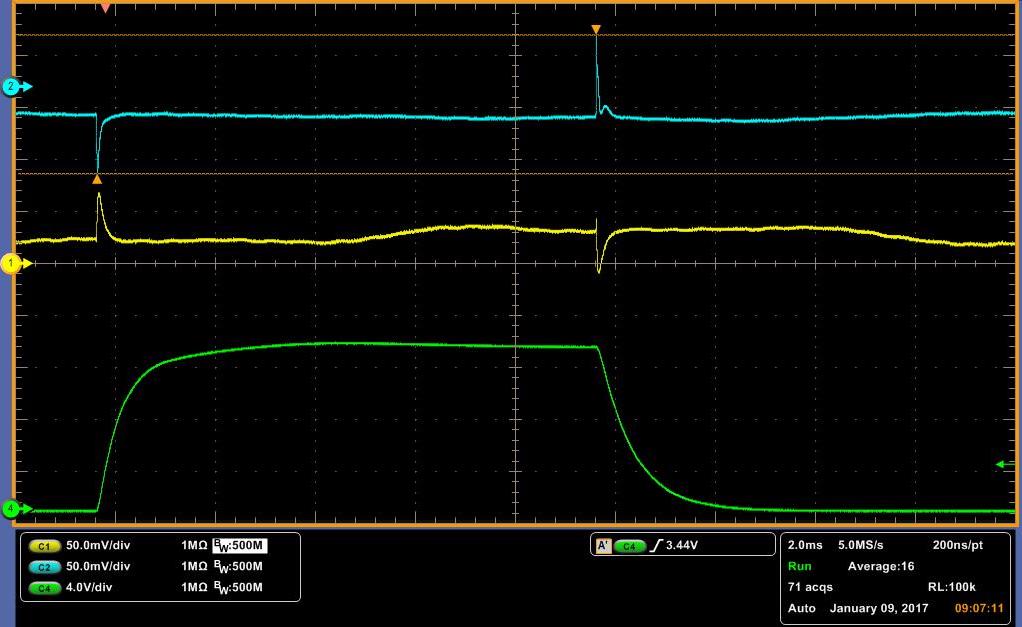}
   \caption{\label{acct}Beam current measurements on the LEBT line. Green: Faraday cup, yellow and blue: ACCT signals.}
\end{figure}

Faraday cups and ACCTs are used to measure and compare the peak beam current at the LEBT line. Box-1 has a locally designed aluminum Faraday cup with an electron suppressor and an inner diameter of 80 mm. Instead of using an aperture at the entrance, the Faraday cup was designed as long as possible, 20 cm, to recapture the back-scattered protons. The Faraday cup in the Box-2 is a simple copper tube with 15 mm aperture to measure spot sizes comparable with the RFQ acceptance. The beam current was measured as 12.5 ${\mu}$A for 200 W RF power at Box-2 (Fig.\ref{acct}). 

As an another current measurement method, two ACCTs were designed and installed in-air at input and output sides of the Box-1. The output signals of the toroids were magnified with an amplifier circuit to measurable levels. The waveforms generated by the proton beam are presented in Fig.\ref{acct}. The beamline calibration of the ACCTs is an ongoing process.

\section{RFQ Beam Matching}
\begin{figure}[h]
   \centering
   \includegraphics[width=19pc]{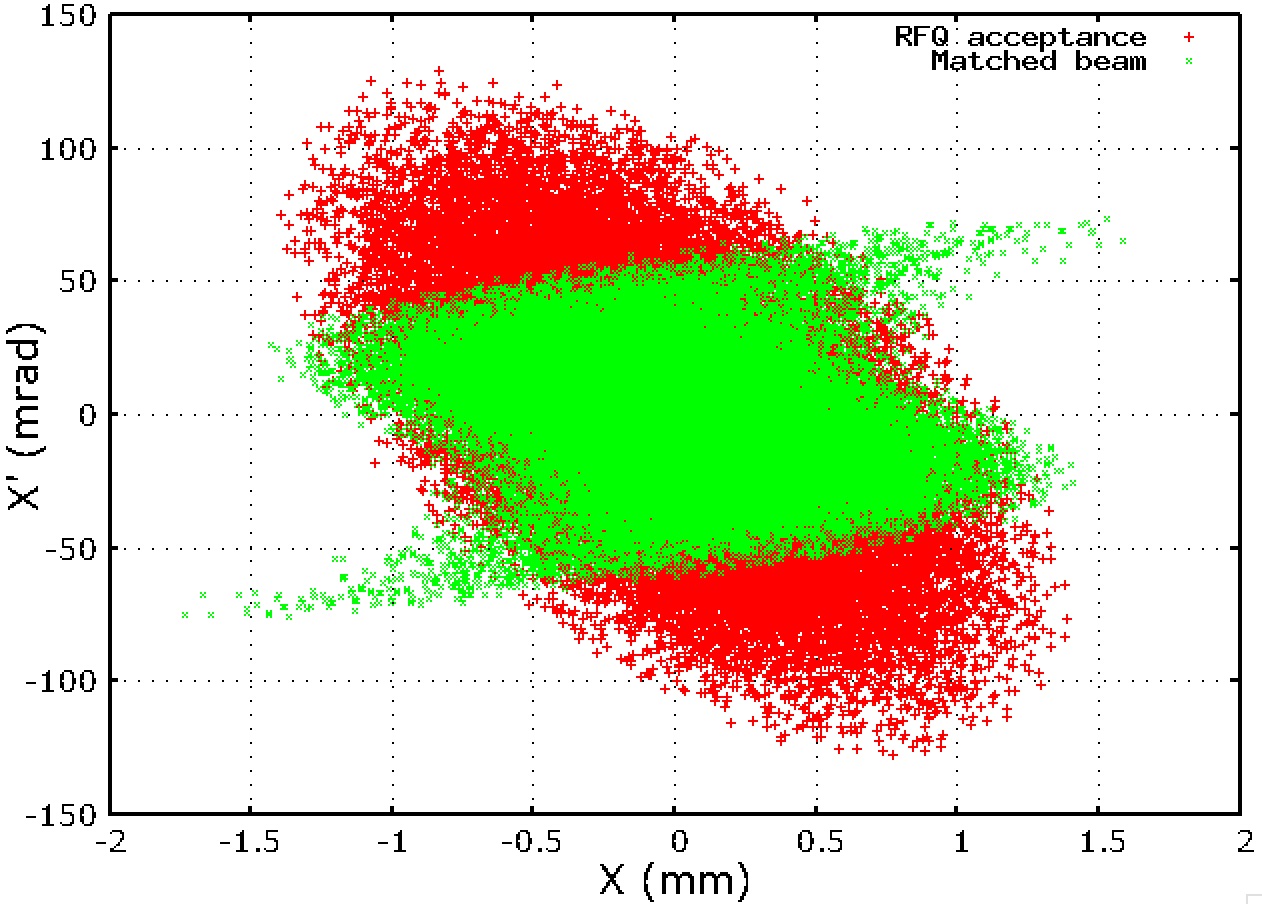}
   \caption{\label{accep}Matched beam on RFQ acceptance (RFQ acceptance in red and matched beam in green).}
\end{figure}

The acceptance of the RFQ is calculated with LIDOS.RFQ \cite{lidos} code and obtained as 0.1 ${\mu}$m (normalized rms). A number of simulations were performed with the TRAVEL \cite{travel} code to estimate the SOL-1 and SOL-2 currents to match the beam to the RFQ acceptance. A method based on emittance measurements was employed to reconstruct the proton beam due to the difficulty in characterizing the ICP plasma. In this method, the extracted proton beam was let to pass through the SOL-1 and its transverse emittance was measured in the Box-1. Then the beam is reconstructed at pepper-pot plane and transmitted through the SOL-2 to match the beam to the RFQ. The proton beam was reconstructed as an ellipse with TRAVEL code using the twiss parameters of the measured phase spaces. These parameters are obtained with a home-built pepper-pot code by analyzing the pepper-pot image. The field maps of the solenoids are generated by the field measurements and used in the TRAVEL code. The beam was matched to the RFQ acceptance (Fig.\ref{accep}) with SOL-1 and SOL-2 currents of 13.00 and 10.16 A, respectively, according 
to results of the TRAVEL simulations. 

Another method for RFQ matching which is required to characterize the ICP plasma and also fully based on IBSimu \cite{ibsimu} and TRAVEL simulations is introduced elsewhere \cite{ipac15s}. Comparison with the results of the latter method will not be meaningful unless the ICP plasma parameters are precisely characterized.

\section{High Power RF Tests}
The RF power requirement of the RFQ is 94.2 kW which corresponds to a measured unloaded quality factor ($Q_{0}$) of 6426 after assembly \cite{ipac16}. A 60 dB RF load, with a maximum power rating of 50 kW, was used for high power RF measurements. The high power RF measurements were performed at the RF PSU, RF transmission line and RFQ cavity.

A hybrid RF power supply was developed locally using 8 solid-state amplifier (SSA) cards and a tetrode RF tube \cite{ipac15t}. According to tests, an RF power of 5.8 kW was reached with 0.01\% duty factor.  An output power of 15.9 kW was measured and a gain of 13.1 dB was calculated with 780.3 W input power. Hence an input power of 4.6 kW is enough to provide the RF power required for the RFQ cavity.

A locally designed and built RF transmission line \cite{EM} was assembled for which an insertion loss of -0.34 dB was measured with a VNA. Thus, as expected, a power reduction of \%7.48 was measured at 16 kW RF power level. An RF power of 47.39 kW was delivered to the end of the line which is close to the acceptable limit for the RF load in use. Beyond this power level, the power level of the RF system was measured with the RFQ cavity itself.

A coaxial loop coupler was designed, built and assembled to feed the RF power to the cavity. A pick-up loop was inserted into the RFQ cavity with -60.8 dB coupling to measure the cavity field. A HPGe detector was used to estimate the inter-vane voltage of the RFQ vanes with the x-ray spectroscopy method. Up to now 23.6 kV inter-vane voltage was reached. The power level of the cavity is gradually increased to provide 60 kV inter-vane voltage for the RFQ cavity with low spark rate.

\section{Outlook}
The required parameters (current, beam size and emittance) are measured to characterize the LEBT beam, except the beam energy which will be obtained by the time-of-flight method in further studies. As the results of the measurements it can be seen that the beam could match to the RFQ with determined configurations. Following the high power RF conditioning of the cavity, the low energy beam will be let to pass the RFQ. The beamline is now fully installed and is being readied for MeV level operations. The first MeV energy proton beam is expected to be available in 2017.

\section{Acknowledgments}
This study is funded by TAEK with a project No. A4.H4.P1. The authors would like to thank A. Lombardi for providing the TRAVEL software.

\end{document}